\documentclass{iopconfser}
\usepackage{url,color}
\usepackage{graphicx} 
\usepackage{amsfonts}
\usepackage{amsmath}
\usepackage{caption}
\usepackage{mathrsfs}
\usepackage{subcaption}
\usepackage[justification=centering]{caption}

\usepackage{doi}

\begin{document}

\title{ Benjamin-Ono dynamics of internal waves with currents }

\author{Lyudmila Ivanova }

\affil{School of Mathematics and Statistics, Technological University Dublin, Grangegorman,
Dublin D07 ADY7, Ireland }

\email{D22127663@mytudublin.ie}

\begin{abstract} 

Internal water waves arise when there is a change in density stratification in a fluid, which may occur in an oceanographical context due to variations in temperature, salinity, or other fluctuations in the equations of state. 
We present a derivation of nonlinear integrable models for the propagation of interfacial internal waves arising between two fluid layers of different densities (at the so called pycnocline). 

We examine the integrable Benjamin-Ono (BO) equation as an internal wave model, incorporating underlying currents by permitting a sheared current in both fluid layers. The BO equation arises for a specific small-amplitude asymptotic regime.
We show that the BO soliton characteristics are strongly affected by the shear current parameters.

\end{abstract}

\section{Introduction}

The study of internal and surface waves has been the focus of extensive research, reflecting their importance for both theoretical fluid dynamics and practical oceanographic applications \cite{BO1,Lan08,ChLee,Bow}. In the equatorial Pacific Ocean, internal waves are strongly influenced by vertical stratification, where a shallow layer of warm, lighter water overlies a much deeper layer of colder, denser water. These layers are separated by sharp gradients in temperature (thermocline) and density (pycnocline), which inhibit vertical mixing and provide a stratified environment that governs internal wave propagation \cite{Boyd,AC,CJ}.

Ocean currents further modulate these wave dynamics. In the equatorial Pacific, the trade winds drive a strong westward surface current, while beneath it the Equatorial Undercurrent (EUC) flows eastward as one of the fastest and most stable subsurface jets. Numerous studies have shown that the interaction between EUC and internal waves plays a central role in large-scale fluid motion, energy transfer, and climate variability \cite{Boyd,AC,CJ,CompelliIvanov1}. To capture such interactions, simplified mathematical models are indispensable.

Perturbative techniques have long been employed to derive reduced models for internal waves, leading to classical equations such as the Korteweg–de Vries (KdV) \cite{KdV} model, intermediate long-wave models (ILW) \cite{ChLee,JoEg}, and the Benjamin–Ono (BO) equation \cite{BO1,BO2} and others. In addition, it is possible to take into account the presence of currents and their impact on the internal wave dynamics. The corresponding KdV, ILW and BO models can be obtained through the Hamiltonian framework of Zakharov \cite{Z68}, represented via the so-called Dirichlet–Neumann operators \cite{CGK}, in both variable and flat-bottom settings, see, for example, \cite{CIM-16, CI-19, CoIv2, CuIv, HIS, Iv17, II25,IMT}. In this study, we investigate the impact of the currents on the solitary internal waves that form between two layers of constant density. We consider a flat surface approximation and an ``infinitely'' deep lower layer which leads to a model, represented by the BO equation \cite{CI19,CuIv,II25}.
This approximation is suitable for modeling equatorial internal waves interacting with EUC, since the top layer in the Pacific ocean is 100-400 meters deep, while the bottom layer is 1000-4000 meters deep \cite{Boyd}.   

The BO equation is particularly notable for its integrability via the inverse scattering transform, which admits exact soliton solutions \cite{FA,KM98,Ma1}.  Solitons are localized, non-dispersive waves that maintain shape and speed even after mutual interactions. Therefore, solitary internal waves form stable geophysical structures that are 
susceptible to observations and analysis.

In this contribution, we investigate the soliton solutions of the BO equation, provide a detailed comparison of their amplitudes and energies, and discuss how these characteristics reflect the flow characteristics of the shear current.

\section{Problem Formulation and Governing Equations}

The equatorial internal wave is represented by a two-dimensional, two-layer incompressible and inviscid fluid model with constant densities, under the assumption that meridional motion is negligible. 
\begin{figure}[h!]
\vspace{-50pt}
    \centering
    \begin{subfigure}{0.48\textwidth}
        \centering
    \includegraphics[width=\textwidth]{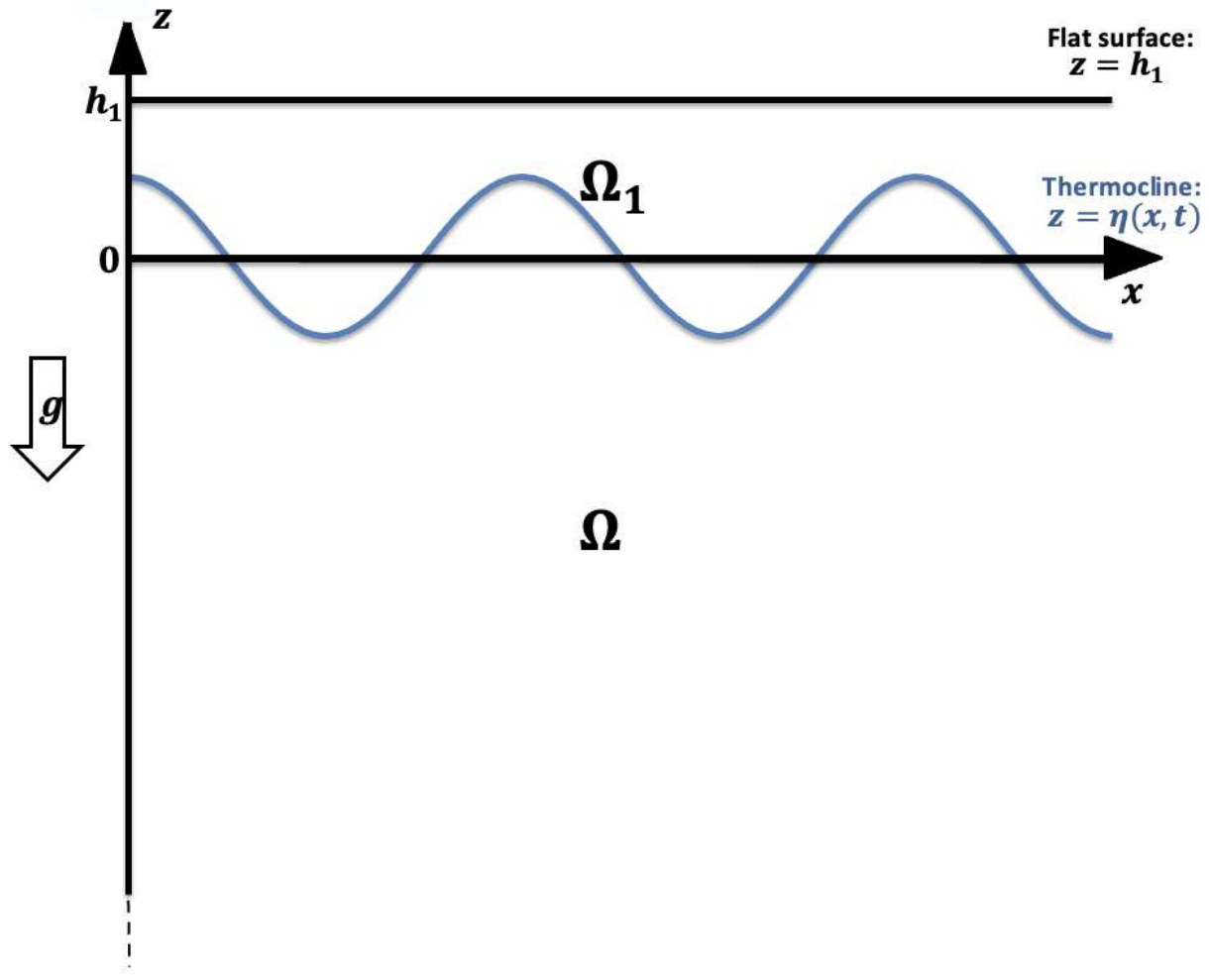}
    \vspace{-70pt}
    \caption{The fluid domain.}\label{fig:fluid}\end{subfigure}
    \hfill
    \begin{subfigure}{0.5\textwidth}
        \centering
    \includegraphics[width=0.95\textwidth]{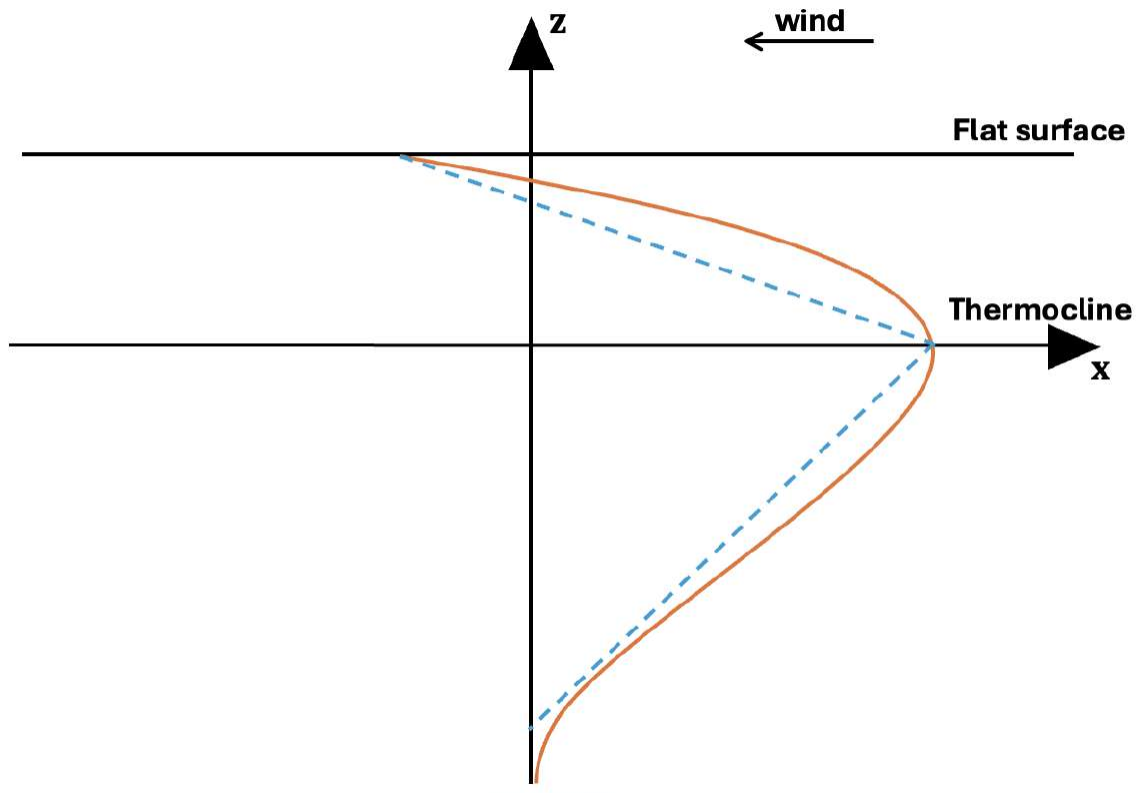}
   \vspace{-22pt}
    \caption{Current profile.}\label{fig:current}
\end{subfigure}
    \caption{The setup.}
    \label{fig:setup}
\end{figure}
The fluid domain is divided into two regions, separated by the interface $z=\eta(x,t),$ which represents the thermocline/pycnocline:
\begin{align}
        \Omega & := \{(x,z)\in \mathbb{R}^{2}:-\infty<z<\eta(x,t)\}, \\
    \Omega_{1}& :=\{(x,z)\in \mathbb{R}^{2}:\eta(x,t)<z<h_{1}\}.
\end{align}
 The upper layer $\Omega_{1}$ has density $\rho_{1}$ and finite depth, bounded above by the flat free surface at $z=h_{1}$. The lower layer $\Omega$ has density $\rho$ with $\rho>\rho_{1}$ and extends infinitely deep, Fig. \ref{fig:setup}(a).

The interface displacement $\eta(x,t)$ represents the vertical elevation of the pycnocline relative to its equilibrium at $z=0$. For analytical convenience, it is assumed that $\eta$ has zero mean:
\begin{equation} \label{eta=0}
    \int_{\mathbb{R}}\eta(x,t)dx = 0,
\end{equation}
for all $t$. The internal wave propagates along the positive $x$-axis, while gravity acts in the negative $z$-direction. This setup ensures stable stratification, with the lighter upper layer overlying the denser lower one, Fig. \ref{fig:setup}(a).

The velocity field is defined as
\begin{equation}
    \mathbf{V}(x,z,t)=(u,w) \,\, \text{in} \, \, \Omega, \quad   \mathbf{V}_{1}(x,z,t)=(u_{1},w_{1}) \,\, \text{in} \, \, \Omega _1
\end{equation}
and the incompressibility condition implies
\begin{equation} \label{eq3}
    u_{x}+w_{z}=0, \quad  u_{1,x}+w_{1,z}=0.
\end{equation}
Equation (\ref{eq3}) motivates the introduction of the stream functions $\psi$ and $\psi_{1}$, defined by
\begin{equation}
    u=\psi_{z},\ w=-\psi_{x} \ \textrm{in} \ \Omega,
\end{equation}
\begin{equation}
    u_{1}=\psi_{1,z}, \ w_{1}=-\psi_{1,x} \ \textrm{in} \ \Omega_{1}.
\end{equation}
To capture the irrotational components of the flow, which may include steady background currents, we introduce the velocity potentials
\begin{equation}
    \varphi\equiv \tilde{\varphi}+\kappa x, \ \varphi_{1}\equiv \tilde{\varphi}_{1}+\kappa_{1}x
\end{equation}
where $\kappa$ and $\kappa_{1}$ denote the constant speeds components in the two layers. The functions $\tilde{\varphi}$ and $\tilde{\varphi}_{1}$ then represent the wave-induced motion.
In the case of constant vorticity, the velocity components in the region $\Omega$ are given by
\begin{equation}
u=\tilde{\varphi}_{x}+\gamma z+\kappa, \ w=\tilde{\varphi}_{z},
\end{equation}
and in the region $\Omega_{1}$ by
\begin{equation}
u_{1}=\tilde{\varphi}_{1,x}+\gamma_{1} z+\kappa_{1}, \ w_{1}=\tilde{\varphi}_{1,z},
\end{equation}
where $\gamma=-w_{x}+u_{z}$ and $\gamma_{1}=-w_{1,x}+u_{1,z}$ are the constant nonzero vorticities, see Fig. \ref{fig:setup}(b). This formulation provides a consistent framework for modeling subsurface oceanic currents, including large-scale phenomena such as the equatorial undercurrent \cite{CIM-16,CI-19,Iv17,CompelliIvanov1,CoIv2,CI19}.

Throughout the analysis, we assume that $\eta$ and $\tilde{\varphi}$ belong to the Schwartz class with respect to the spatial variable $x$, for each fixed $(z,t)$. Similarly, $\tilde{\varphi}_{1}$ is assumed to belong to the Schwartz class with respect to $x$ and $z$, for each fixed $t$. These assumptions guarantee smoothness and rapid decay at infinity, which are essential for the derivations. In particular, they imply that internal waves vanish at infinity:
\begin{equation}
    \lim_{|x|\to \pm\infty}\eta(x,t)=\lim_{|x|\to \pm\infty} \tilde{\varphi}(x,z,t)=\lim_{|x|\to \pm\infty}\tilde{\varphi}_{1}(x,z,t)=0.
\end{equation}
Moreover, we require that no wave motion persists at infinite depth, which yields
\begin{equation}
    \lim _{z\to -\infty}\tilde{\varphi}(x,z,t)=0.
\end{equation}

The Euler equations for each fluid layer are

\begin{equation}
     \mathbf{V}_{t}+(\mathbf{V}\cdot \nabla)\mathbf{V} = -\frac{1}{\rho}\nabla p+\mathbf{g},
\end{equation}
\begin{equation}
     \mathbf{V}_{1,t}+(\mathbf{V}_{1}\cdot \nabla)\mathbf{V}_{1} = -\frac{1}{\rho_{1}}\nabla p_{1}+\mathbf{g}.
\end{equation}
Here $p$ and $p_{1}$ are the pressures in the two layers, and $\mathbf{g}=(0,-g)$ is the gravitational acceleration.

Reformulating in terms of the introduced variables makes the pressure gradients explicit. The dynamic boundary condition $p=p_{1}$ yields
\begin{equation}
\label{eq13}
    (\rho\tilde{\varphi}_{t}-\rho_{1}\tilde{\varphi}_{1,t})_{i}=\frac{\rho_{1}}{2}|\nabla\psi_{1}|^{2}_{i}-\frac{\rho}{2}|\nabla\psi|_{i}^{2}+(\rho\gamma-\rho_{1}\gamma_{1})\chi+(\rho_{1}-\rho)g\eta,
\end{equation}
a Bernoulli-type relation. The subscript $i$ indicates evaluation at the interface $z=\eta(x,t)$. The stream function $\chi$ is
\begin{equation}
    \chi(x,t):=(\psi)_{i}=(\psi_{1})_{i}=-\int_{-\infty}^{x}\eta_{t}(x',t)dx'=-\partial_{x}^{-1}\eta_{t}.
\end{equation}

Equation (\ref{eq13}) governs the evolution of $\xi:=\rho\tilde{\varphi}-\rho_{1}\tilde{\varphi}_{1}$, a momentum-type variable in the Hamiltonian setting. 

The interface position $\eta(x,t)$ satisfies the kinematic condition
\begin{equation}
    \eta_{t}=w-\eta_{x}u=w_{1}-\eta_{x}u_{1},
\end{equation}
or
\begin{equation}
\label{eq16}
    \eta_{t}=(\tilde{\varphi}_{z})_{i}-\eta_{x}((\tilde{\varphi}_{x})_{i}+\gamma\eta+\kappa)=(\tilde{\varphi}_{1,z})_{i}-\eta_{x}((\tilde{\varphi}_{1,x})_{i}+\gamma_{1}\eta+\kappa_{1}).
\end{equation}

At the upper (flat) surface, the kinematic boundary condition enforces vanishing of the vertical velocity, 
\begin{equation}
    (\tilde{\varphi}_{1}(x,h_{1},t))_{z}=0.
\end{equation}

\section{Long-wave scaling and the Benjamin-Ono model}

The governing equations (\ref{eq13}) and (\ref{eq16}) can be reformulated in a dimensionless form, which facilitates the analysis of the underlying physical processes and highlights the relative importance of different terms. In the case of fluid motion within a two-layer medium, the commonly employed dimensionless scaling parameters are 
\begin{equation}
    \varepsilon:=\frac{a}{h_{1}},\quad    \delta:=\frac{h_{1}}{\lambda} \ll 1,
\end{equation}
where $a$ denotes the characteristic amplitude of the internal waves $\eta(x,t)$, $h_{1}$ is the depth of the upper fluid layer, and $\lambda$ is the characteristic wavelength. $\varepsilon$ is used to characterize the weak nonlinearity of the waves, and $\delta$ defines the long-wave regime. By employing these parameters, the terms in the governing equations are systematically ordered, and simplified models are derived that capture the essential dynamics of weakly nonlinear, long internal waves in a two-layer fluid.

A key assumption in this study is that $\varepsilon$ and $\delta$ are of the same order of magnitude, i.e. $\mathcal{O}(\varepsilon)=\mathcal{O}(\delta)$. The variables are scaled as $\eta\sim\mathcal{O}(\delta)$ and $\partial\sim\mathcal{O}(\delta)$; since the differential operators $\partial$ and $D$ are related by 
\begin{equation}
    D:=-i\partial_{x},
\end{equation}
with the wavenumber $k:=\frac{2\pi}{\lambda}$ appearing as the eigenvalue (or Fourier multiplier) of $D$ for monochromatic waves of the form $e^{ikx}$, it follows that $\partial\sim \mathcal{O}(\delta)$.

For the lower layer, assumed to be infinitely deep, one has $\mathcal{O}(hk)\gg 1$. In fact, what is assumed is $|\tanh (hk)|=1$ then the operator $\tanh(h D) =\text{sign} (D).$ The operator $\text{sign} (D)$ is related to the Hilbert transform, $\mathcal{H}=-i\, \text{sign} (D),$
$$  \mathcal{H}\{f\} (x) := \mathrm{P.V.}\frac{1}{\pi}\int_{-\infty}^{\infty}\frac{f(x')dx'}{x-x'},$$
therefore $\partial_x  \mathcal{H}=D \, \text{sign} (D)=|D|, $  an operator with a Fourier multiplier $|k|.$

In contrast, for the thin upper layer $\mathcal{O}(h_{1}k)\sim\delta$, which reflects the long-wave scaling, and consequently $\mathcal{O}\big({\frac{h_{1}}{h}}\big)\ll \delta $. 

All physical parameters, $h_{1},\kappa, \rho, \rho_{1}, \gamma, \gamma_{1}$, and $g$, (in non-dimensional form) are assumed to be of order unity, $\mathcal{O}(1)$ meaning that they do not introduce additional small or large scales. This ensures that, after nondimensionalization, the governing equations depend only on the dimensionless variables and the long-wave parameter $\delta$.

Based on the above approximations, a nonlinear equation for $\eta(x,t)$ is derived by employing the Hamiltonian approach through Dirichlet–Neumann operators, and is given by \cite{CI19,CuIv}

\begin{equation} 
\label{eq20}
\eta_{t}+c\eta_{x}+\delta\mathscr{A}\eta\eta_{x}-\delta\mathscr{B}|D|\eta_{x}=0.
\end{equation}
Here, the constant $\mathscr{A}$ is defined as
\begin{equation}
        \mathscr{A}:=\frac{-3\rho_{1}(c-\kappa)^{2}+3\rho_{1}\gamma_{1}h_{1}(c-\kappa)+h_{1}^{2}(\rho\gamma^{2}-\rho_{1}\gamma_{1}^{2})}{h_{1}(2\rho_{1}(c-\kappa)+h_{1}\Gamma)},
    \end{equation}
with $\Gamma=\rho\gamma-\rho_{1}\gamma_{1}$ and the constant $\mathscr{B}$ is defined as
\begin{equation}
        \mathscr{B}:=\frac{\rho h_{1}(c-\kappa)^{2}}{2\rho_{1}(c-\kappa)+h_{1}\Gamma}.
\end{equation}
The equation in (\ref{eq20}), known as the Benjamin–Ono equation, is an integrable nonlinear equation whose solutions can be obtained through the inverse scattering method. Its integrability ensures an infinite number of conserved quantities and allows for the exact construction of multi-soliton solutions, making it fundamental in the theory of internal waves in deep stratified fluids. From this equation, it can also be seen that the constant term for an infinitely deep lower layer is $|D|$.

The wave speed $c$ in the leading order ($\delta \to 0$) has the constant value
\begin{equation}
    c-\kappa=-\frac{h_{1}}{2\rho_{1}}\Gamma\pm\frac{1}{2}\sqrt{\frac{h_{1}^{2}}{\rho_{1}^{2}}\Gamma^{2}+4\frac{h_{1}}{\rho_{1}}g(\rho-\rho_{1})},
\end{equation}
with the two signs corresponding to right- and left-propagating waves. These velocities are the same for intermediate-long waves when the lower layer has a finite depth, with   $\mathcal{O}\big({\frac{h_{1}}{h}}\big) = \delta \ll 1,  $ see for example \cite{CuIv}.

It should be noted that the nonlinear term in \eqref{eq20} has a coefficient $\mathscr{A}$. If $\mathscr{A}$ vanishes - which occurs, for example, when $\gamma_{1}^{2}>\frac{\rho_{1}}{4\rho}\gamma^{2}$ — the resulting equation for $\eta(x,t)$ is linear, therefore in this case there is no so soliton solution.

The Benjamin–Ono equation in the irrotational case, where $\gamma=\gamma_{1}=\kappa=0$, takes the form
\begin{equation} \label{BO1}
    \eta_{t}+c_0\eta_{x}-\frac{1}{2}\delta\frac{\rho h_{1}c_0}{\rho_{1}}|D|\eta_{x}-\frac{3}{2}\delta\frac{c_0}{h_{1}}\eta\eta_{x}=0.
\end{equation}
Moreover, we obtain the following expression for $c_0$:
\begin{equation}\label{c0}
    c_0=\pm\sqrt{\frac{h_{1}}{\rho_{1}}g(\rho-\rho_{1})}.
\end{equation}

We notice that the evolution equation \eqref{eq20} depends only on $\eta$-derivatives of ``positive'' order, therefore we can relax the condition \eqref{eta=0} to
\begin{equation} \label{eta=const}
    \int_{\mathbb{R}}\eta(x,t)dx = \text{constant}.
\end{equation}
This will not change the average depth of the thermocline, since the increase of the average depth would be 
$(\int_{\mathbb{R}}\eta(x,t)dx) /( \int_{\mathbb{R}} dx) =0,$ as far as the integral in the denominator is infinity.

\section{Soliton Properties and Comparisons}

The Benjamin–Ono (BO) equation (\ref{eq20}) admits a one-soliton solution, which preserves its shape due to a balance between nonlinearity and dispersion. In the parameterization used here it can be written as

\begin{equation} 
\label{eq27}
    \eta(x,t)=\frac{4\mathscr{B}}{\mathscr{A}}\cdot\frac{K}{1+K^{2}[x-x_{0}-(c+\delta\mathscr{B}K)t]^{2}},
\end{equation}
where $x_{0}$ (the initial position of the crest) and $K$ are the soliton parameters.

The soliton amplitude is $\eta^{*}:=\frac{4\mathcal{B}}{\mathscr{A}}K$, while the propagation speed in this scaling is $c+\delta\mathscr{B}K$. Thus both the soliton amplitude and the soliton speed depend linearly on the soliton parameter $K$. 
The scale factor $\delta$ only indicates that the soliton parameter $K$ is of order $\delta \ll 1$ (then, of course $\eta$ and $\eta^{*}$ are of order $\delta$ as well) and that for physically relevant situations $K\ll 1.$ 
The one–soliton solution of the Benjamin–Ono equation possesses an infinite hierarchy of conserved quantities. The first in this hierarchy is the mass, defined by
\begin{equation}
\label{eq28}
    M=\int_{-\infty}^{\infty}\eta(x,t)dx=\frac{4\pi\mathscr{B}}{\mathscr{A}}.
\end{equation}
This expression depends solely on the ratio $\mathscr{B}/\mathscr{A}$ and is independent of the soliton parameter $K$. In particular, the mass attains the same value for all members of the one–soliton family, regardless of their spatial scale or amplitude.  Formally, one can take the (unphysical) limit $K \to \infty $ to obtain a soliton with the shape of the Dirac delta-function, that propagates also with an infinite speed.

Thus, the different one–soliton profiles may possess the same mass, so this invariant by itself is insufficient to uniquely identify a given solitary–wave solution. Accordingly, higher-order conserved quantities must be considered to distinguish between solitons with differing structural characteristics.

To obtain a more refined characterization, one considers the next conserved quantity, which is usually associated to the  energy of the soliton, 
\begin{equation}
\label{eq29}
    E=\frac{1}{2}\int_{-\infty}^{\infty}\eta(x,t)^{2}dx=\frac{4\pi\mathscr{B}^{2}}{\mathscr{A}^{2}}K.
\end{equation} 
In contrast to the mass, the energy depends explicitly on the soliton parameter $K$, which determines the speed and the amplitude of the soliton. Therefore, the energy can distinguish between solitons that have the same mass but different characteristics. In this sense, it provides more information about the soliton behavior, and it is particularly useful when comparing solitary-wave solutions of the Benjamin–Ono equation. In \cite{Tal} for example, the energy of the soliton has been used to characterize its propagation over a bottom threshold between two depths.

\begin{figure}[h!]
    \centering
\includegraphics[width=0.5\linewidth]{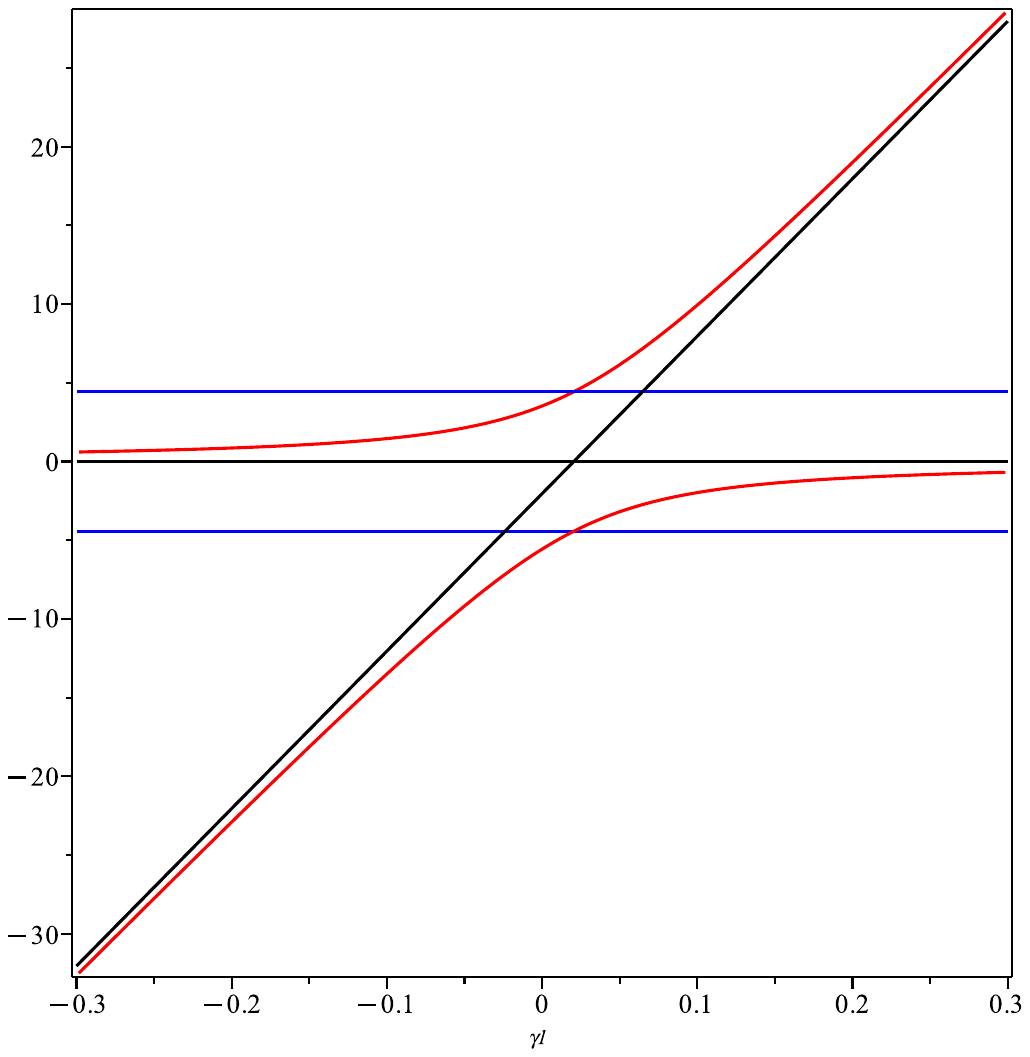}
\vspace{-70pt}
    \caption{$c_{+}$ and $c_{-}$ (red) versus the vorticity in the upper layer $\gamma_1$ (the other shear parameter is $\gamma=0.02$).  $\pm |c_{0}|$ (blue) are the zero-shear values. The black line is the asymptote $c_{\pm}\simeq \gamma_1 h $ of the growing branch of both speeds, the other branch of the speeds (propagating against the current) decays to zero.  }
    \label{fig:2}
\end{figure}

To examine the influence of background shear, we compare the general Benjamin–Ono model (\ref{eq20}) with its irrotational limit \eqref{BO1} with $\gamma=\gamma_{1}=0.$ In the absence of shear, the wave speeds of the left- and the right-running wave \eqref{c0} are of the same magnitude and equal to $\pm |c_{0}|$. When shear is introduced, this symmetry is broken and the speeds split into two branches $c_{\pm}(\Gamma)$, where $\Gamma=\rho\gamma-\rho_{1}\gamma_{1}$ is the effective shear parameter. As illustrated in Figure \ref{fig:2}, the shear increases the speed for waves traveling in the direction of the background current and slows down waves, traveling against it. Moreover, for large values of $|\gamma_1|$ (while $\gamma$ is fixed) the asymptotic behavior is $c_{\pm}\simeq \gamma_1 h,$ this corresponds to the line in black color on Fig. \ref{fig:2}. 


Let us now investigate the influence of shear on the nonlinearity and dispersion coefficients $\mathscr{A}$ and $\mathscr{B}.$ For simplicity, in what follows, we examine only the waves running to the right with wave speed $c_+$. Physically, these are the waves that travel to the East, which occur in the equatorial Pacific Ocean.

\begin{figure}[h!]
    \centering
    \begin{subfigure}{0.48\textwidth}
        \centering
    \includegraphics[width=\textwidth]{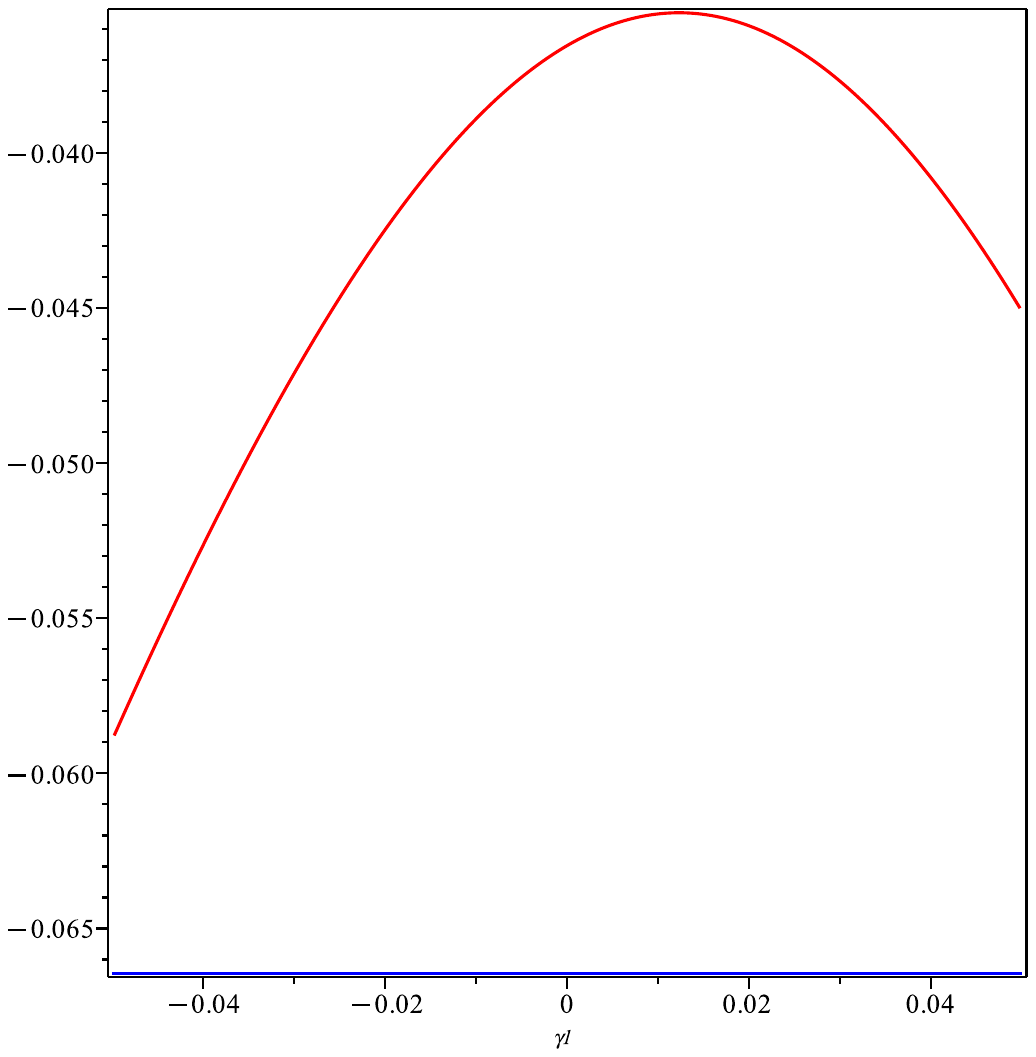}
    \vspace{-70pt}
    \caption{Coefficient $\mathscr{A}$(red) versus its irrotational limit $\mathscr{A}_{0}$(blue line).}\label{fig:A}\end{subfigure}
    \hfill
    \begin{subfigure}{0.48\textwidth}
        \centering
    \includegraphics[width=\textwidth]{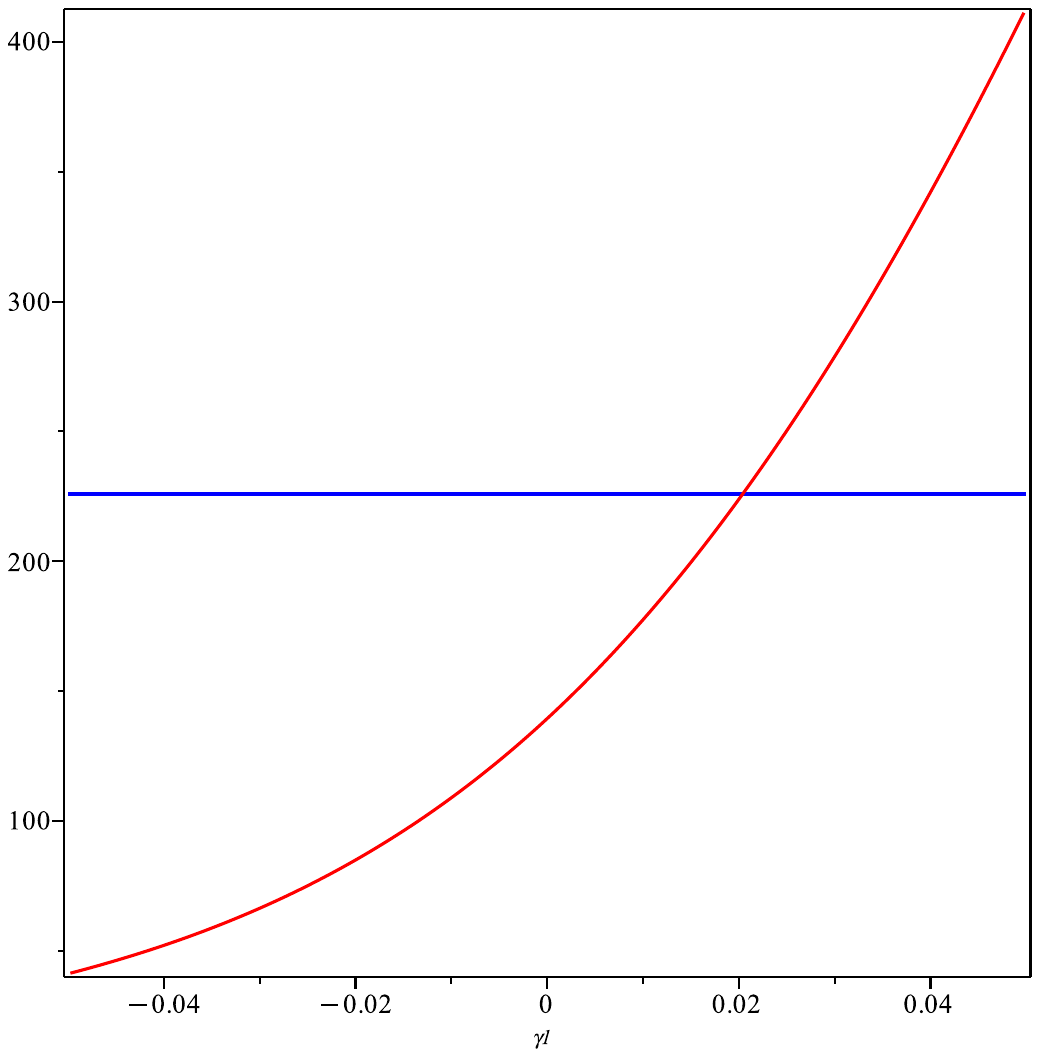}
    \vspace{-70pt}
    \caption{Coefficient $\mathscr{B}$ (red) versus its irrotational limit $\mathscr{B}_{0}$(blue line).}\label{fig:B}
\end{subfigure}
    \caption{Coefficients $\mathscr{A}$ and $\mathscr{B}$ for $\gamma=0.02$ and varying $\gamma_{1}$, compared with the irrotational case $\mathscr{A}_{0}$	
  and $\mathscr{B}_{0}$ for $\gamma=\gamma_{1}=0$.}
    \label{fig:AB}
\end{figure}

In Figures \ref{fig:AB}(a) and \ref{fig:AB}(b), the dependence of $\mathscr{A}$ and $\mathscr{B}$ on $\gamma_1$ is shown for fixed $\gamma=0.02.$ The irrotational limits ($\gamma=\gamma_{1}=0$), denoted by $\mathscr{A}_{0}=-\frac{3c_0}{2h_1} $ and $\mathscr{B}_{0}=\frac{\rho h_{1}c_0}{2\rho_{1}}$, are included as a reference.
 
As $\gamma_{1}$ increases, non-monotonic behavior is observed for the nonlinearity coefficient $\mathscr{A}$: it is seen to increase initially, reach a maximum, and subsequently decrease. In contrast, the dispersion coefficient $\mathscr{B}$ is found to increase monotonically with $\gamma_{1}$, indicating that the dispersive contribution is strengthened in the presence of background shear. As a result, shear changes the balance between nonlinearity and dispersion, which leads to noticeable differences in the solitary‐wave profile.


For the Benjamin–Ono equation, the energy of a one–soliton solution is given by (\ref{eq29}) where $K$ is the soliton parameter. In the irrotational case we similarly have
\begin{equation}
    E_{0}=\frac{4\pi\mathscr{B}_{0}^{2}}{\mathscr{A}_{0}^{2}}K_{0},
\end{equation}
where $\mathscr{A}_{0}$ and $\mathscr{B}_{0}$ denote the corresponding coefficients for $\gamma=\gamma_{1}=0$. To ensure a proper comparison, we impose the condition of equal energy, $E=E_{0}$, which yields the following relation between the parameters $K$ and $K_{0}$:
\begin{equation}
    \frac{K}{K_{0}}=\Big(\frac{\mathscr{A}\mathscr{B}_{0}}{\mathscr{A}_{0}\mathscr{B}}\Big)^{2}.
\end{equation}
This way, the two solitary waves of equal energy are compared, ensuring that any observed differences in their profiles and amplitudes are due to the presence of background shear.

\begin{figure}[h!]
    \centering
\includegraphics[width=0.5\linewidth]{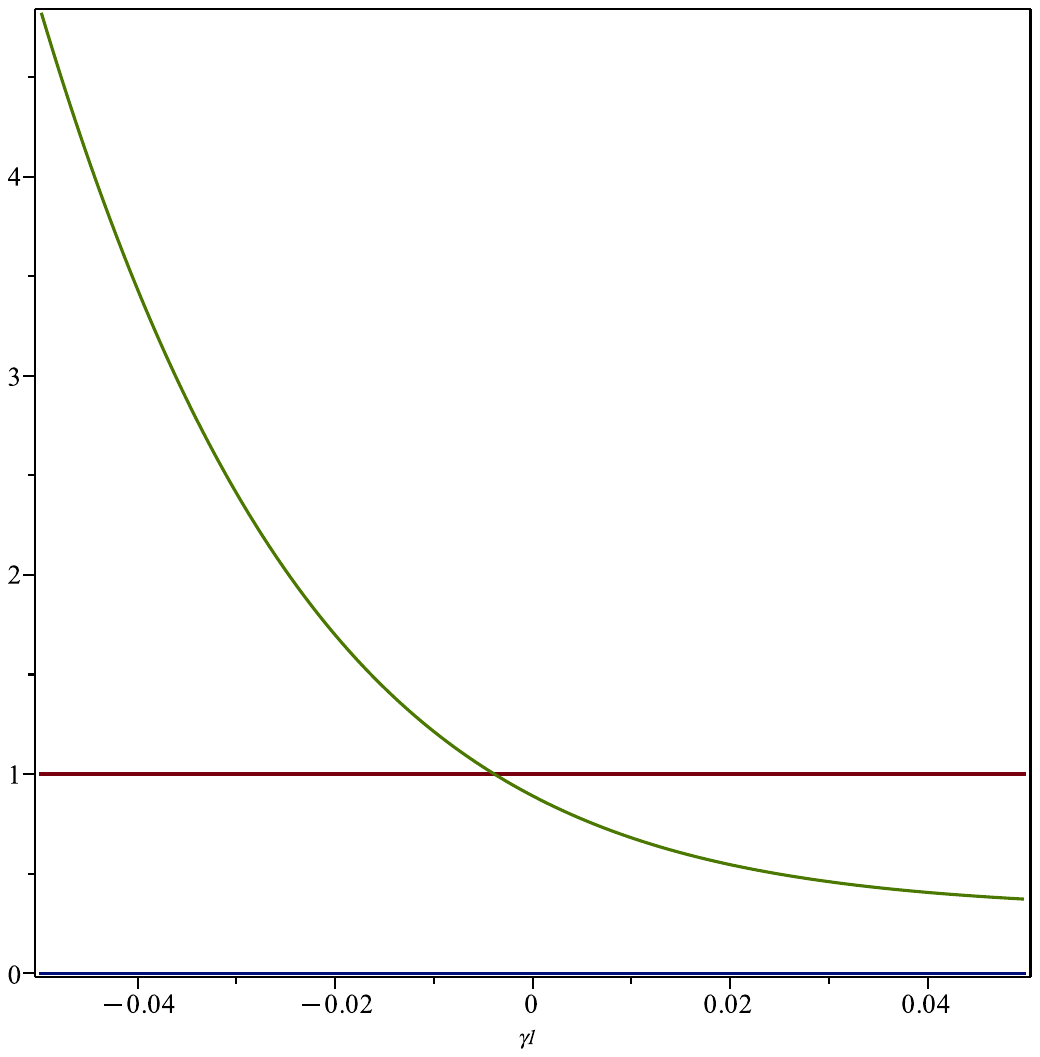}
\vspace{-70pt}
    \caption{Amplitude ratio $R=\eta_s^*/\eta_{0}^*$ at equal energy as a function of the shear parameter $\gamma_{1}$ for fixed $\gamma=0.02$. The horizontal line corresponds to the irrotational value $R=1$.}
    \label{fig:4}
\end{figure}

In order to examine how the background current influences the wave shape, we introduce the amplitude ratio
\begin{equation}
    R:=\frac{\eta_{s}^{*}}{\eta_{0}^{*}}=\frac{\mathscr{A}\mathscr{B}_{0}}{\mathscr{A}_{0}\mathscr{B}},
\end{equation}
where $\eta_{s}^{*}$ is the soliton amplitude in the presence of current, and $\eta_{0}^{*}$ is the amplitude without it. 
Figure \ref{fig:4} shows how $R$ depends on the parameter $\gamma_{1}$ for fixed $\gamma=0.02$. The curve decreases monotonically, demonstrating that stronger shear consistently reduces the value of $R$. Given that in the Pacific Ocean realistically $\gamma_1<0,$ as shown on Fig. \ref{fig:setup}(b), this means that, under the equal-energy condition, the current amplifies the amplitude of the solitary waves, when compared to those in the irrotational case 
($\gamma=\gamma_{1}=0$).


\section{Conclusion}
In this study, we examine internal waves in a two-layer system with a thin upper layer and an infinitely deep lower layer, using an asymptotic approximation that leads to the Benjamin–Ono equation. The results show that background shear noticeably alters the behavior of the model compared to the irrotational case ($\gamma=\gamma_{1}=0$). The wave speeds change in the presence of (constant) vorticity, with waves traveling faster along the shear and slower against it. The coefficients of the nonlinear and dispersive terms of the equation also change and depend significantly on the vorticity parameters, which affects how the waves evolve. Because of this, the solitary-wave profiles and, in particular, the soliton amplitudes change when the solitons are compared to equal energy solitons at irrotational flow. Future work may involve comparing these findings with other reduced models such as the KdV and Intermediate Long Wave equations, which would help for the broader and more comprehensive understanding of how shear influences internal waves.

\subsection*{Acknowledgments} This publication has emanated from research conducted with the financial support of Taighde \' Eireann – Research Ireland under Grant number 21/FFP-A/9150.

\end{document}